\documentclass[preprint, superscriptaddress, preprintnumbers, amsmath, amssymb]{revtex4}

\allowdisplaybreaks
\allowdisplaybreaks[4]
\usepackage{graphicx,latexsym}
\usepackage{multirow,array,booktabs}
\usepackage{subfigure}
\usepackage[figuresright]{rotating}
\usepackage{color}
\usepackage{CJK}
\usepackage{dcolumn}
\usepackage{bm}
\usepackage[dvipdfm,
            pdfstartview=FitH,
            CJKbookmarks=true,
            bookmarksnumbered=true,
            bookmarksopen=true,
            colorlinks=true,
            pdfborder=001,
            linkcolor=blue,
            anchorcolor=blue,
            citecolor=blue
            ]{hyperref}

\begin{document}


\title{Explore nuclear multiple chirality in $A\sim60$ mass
region within covariant density functional theory}

\author{J. Peng}\email{jpeng@bnu.edu.cn}
\affiliation{Department of Physics, Beijing Normal University,
Beijing 100875, China}

\author{Q. B. Chen}
\affiliation{Physik-Department, Technische Universit\"{a}t
M\"{u}nchen, D-85747 Garching, Germany}

\begin{abstract}

The nuclear multiple chirality in $A\sim60$ mass region is explored by the adiabatic and
configuration-fixed constrained covariant density functional theory for cobalt isotopes.
The potential-energy curves and triaxiality parameters $\gamma$ as functions of the
deformation parameter $\beta$ in $^{54,56,57,58,59,60}$Co are obtained.
It is found that there are high-$j$ particle(s) and hole(s) configurations with
prominent triaxially deformed shapes in these isotopes.
This points towards that the existence of chirality or multiple chirality
in $A\sim60$ mass region are highly anticipated.

\end{abstract}


\maketitle

The study of rotational motion has been at the forefront of nuclear
structure physics for many decades. In particular, the chiral
rotation that originally predicted in
1997~\cite{Frauendorf1997Nucl.Phys.A131} is an exotic rotational
phenomenon in triaxially deformed nucleus with high-$j$ valence
particle(s) and high-$j$ valence hole(s). Subsequently, the
prediction of multiple chiral doublets
(M$\chi$D)~\cite{Meng2006Phys.Rev.C37303} provides more and more
impetus for this research area.

Experimentally, the chiral doublet bands were first observed in
$N=75$ isotones~\cite{Starosta2001PRL} in 2001. From then on, the
manifestations of chirality have been extensively confirmed in the
mass regions of $A\sim 80, 100, 130$, and 190
~\cite{J.Meng2010JPG,J.Meng2014IJMPE,Bark2014IJMPE,J.Meng2016PS,Raduta2016PPNP,Starosta2017PS}.
More than 50 chiral candidates spread over 46 nuclei have been
reported in these four mass regions so far. For details, see very
recent data tables of chiral doublets bands~\cite{Xiong2018arXiv}.
On the aspect of M$\chi$D, the first experimental evidence was
reported in $^{133}$Ce~\cite{Ayangeakaa2013PRL}. Subsequently, a
novel type of M$\chi$D with identical configuration, predicted in
Refs.~\cite{Droste2009EPJA,Q.B.Chen2010PRC,Hamamoto2013PRC}, was
observed in $^{103}$Rh~\cite{Kuti2014PRL}. Later on, the M$\chi$D
with octupole correlations has been identified in
$^{78}$Br~\cite{C.Liu2016PRL}. Very recently, M$\chi$D were further reported in the even-even nucleus in $^{136}$Nd~\cite{C.M.Petrache2018PRC41304,Q.B.Chen2018PLB744}
and in the $A\sim190$ mass region in $^{195}$Tl~\cite{T.Roy2018PLB768}.

Theoretically, the chiral symmetry breaking was firstly predicted in
the particle-rotor model (PRM) and tilted axis cranking (TAC)
approach~\cite{Frauendorf1997Nucl.Phys.A131}. Later on, numerous
efforts were devoted to the development of PRM~\cite{C.M.Petrache2018PRC41304,Q.B.Chen2018PLB744, Zhang2016Chin.Phys.C24102, Peng2003Phys.Rev.C44324,
Koike2004PRL, S.Y.Wang2007PRC, S.Q.Zhang2007PRC, B.Qi2009PRC,
B.Qi2009PLB, B.Qi2011PRC, Lieder2014PRL} and TAC
methods~\cite{Dimitrov2000PRL, Olbratowski2004PRL,
Olbratowski2006PRC} to describe chiral rotation in atomic
nuclei. Meanwhile, the TAC plus random-phase approximation
(RPA)~\cite{Almehed2011PRC}, the collective Hamiltonian
method~\cite{Q.B.Chen2013PRC,Q.B.Chen2016PRC}, the interacting
boson-fermion-fermion model~\cite{Brant2008PRC}, and the angular
momentum projection (AMP)
method~\cite{Bhat2012PLB,Bhat2014NPA,F.Q.Chen20017PRC,Shimada2018PRC}
were also developed to study chiral doublet bands and yielded lots
of successes.

The covariant density functional theory (CDFT) takes Lorentz
symmetry into account in a self-consistent way and has received wide
attention due to its successful descriptions for a large number of
nuclear phenomena~\cite{Reinhard1996Rep.Prog.Phys.,
Ring1996Prog.Part.Nucl.Phys., Serot1997Int.J.Mod.Phys.E,
Vretenar2005PhysRep, J.Meng2006PPNP}. Especially for nuclear
rotation, the cranking CDFT was first used to investigate the
superdeformed bands~\cite{Konig1993PRL}. Subsequently, the tilted
axis cranking CDFT~\cite{Madokoro2000PRC,Peng2008Phys.Rev.C24313,Zhao2011Phys.Lett.B181}
has successfully provided the fully self-consistent and microscopic
investigation for
magnetic~\cite{Madokoro2000PRC,Peng2008Phys.Rev.C24313,Zhao2011Phys.Lett.B181,
Steppenbeck2012Phys.Rev.C44316,Yu2012Phys.Rev.C24318}
and antimagnetic
rotations~\cite{Zhao2012Phys.Rev.C54310,P.Zhang2014PRC,Peng2015Phys.Rev.C44329}.
Recently, the three-dimensional cranking CDFT was
established~\cite{Zhao2017PLB} and applied for the chiral
bands in $^{106}$Rh~\cite{Zhao2017PLB} and
$^{136}$Nd~\cite{Petrache2018PRC}. Based on constrained triaxial
CDFT calculations, M$\chi$D phenomenon, namely more than one pair of
chiral doublet bands in one single nucleus, was suggested for
$^{106}$Rh in 2006~\cite{Meng2006Phys.Rev.C37303}. Later on, the
existence of M$\chi$D phenomenon was also suggested in
$^{104,105,106,108,110}$Rh~\cite{Peng2008Phys.Rev.C24309,Yao2009PRC067302,J.Li2011PRC},
$^{107}$Ag~\cite{B.Qi2013PRC}, and
${^{125,129,131}}$Cs~\cite{J.Li2018PRC} based on the triaxial
deformations of the local minima and the corresponding high$-j$
particle(s) and hole(s) configurations that obtained by constrained
CDFT calculations.

As mentioned above, lots of bands with chiral rotation have been
identified in the $A \geqslant 80$ mass regions. It is natural to
examine the existence of chirality in the nuclear system with
lighter mass. As a first try, we pay attentions on the $A\sim60$
mass region and select the cobalt isotopes as possible candidates in
this paper. Following the same procedures outlined in
Refs.~\cite{Meng2006Phys.Rev.C37303,Peng2008Phys.Rev.C24309}, the
possible configurations as well as their deformation parameters in
the cobalt isotopes $^{54,56,57,58,59,60}$Co will be examined by the
adiabatic and configuration-fixed constrained triaxial CDFT.
The reason for the choice of Co isotopes is that the proton number of
Co is odd with the proton already playing a role of high-$j$ ($f_{7/2}$) hole
according to the Nilsson diagram~\cite{Nilsson1955MFMDVS,Ring1980book}.
To establish the chiral doublet bands, one needs only search for the case that
neutron plays a role of high-$j$ particle. This makes the chiral configuration
become more energy favor, and easier been populated in the experiment.

The detailed formalism and numerical techniques of the adiabatic and
configuration-fixed constrained CDFT calculation adopted in this
work can be seen in Refs.~\cite{Meng2006Phys.Rev.C37303,
Peng2008Phys.Rev.C24309} and references therein. In the
calculations, the point-coupling density functional
PC-PK1~\cite{Zhao2010Phys.Rev.C54319} is employed, while the pairing
correlations are neglected for simplicity. The Dirac equation is
solved in a set of three dimensional harmonic oscillator basis. By
increasing the number of major shells from 12 to 14, the total
energy changes less than 0.1$\%$ for the ground state of $^{54}$Co.
Therefore, a basis of 12 major oscillator shells is adopted. In
order to obtain the triaxiality of the local minima on the potential
energy curve, the constrained calculations with $\left<
\hat{Q}_{20}^2+2\hat{Q}_{22}^2 \right>$, i.e., the $\beta^2$, are
carried out. During the $\beta^2$-constrained calculations, triaxial
deformation is automatically obtained by minimizing the energy. In
order to understand configurations easily, following
Ref.~\cite{Peng2008Phys.Rev.C24313}, the wave functions in the
Cartesian basis $\left< {n_xn_yn_zm_s}\right>$ are transformed to a
spherical basis with the quantum number $\left< {nljm}\right>$.


\begin{figure*}
\includegraphics[width=12 cm]{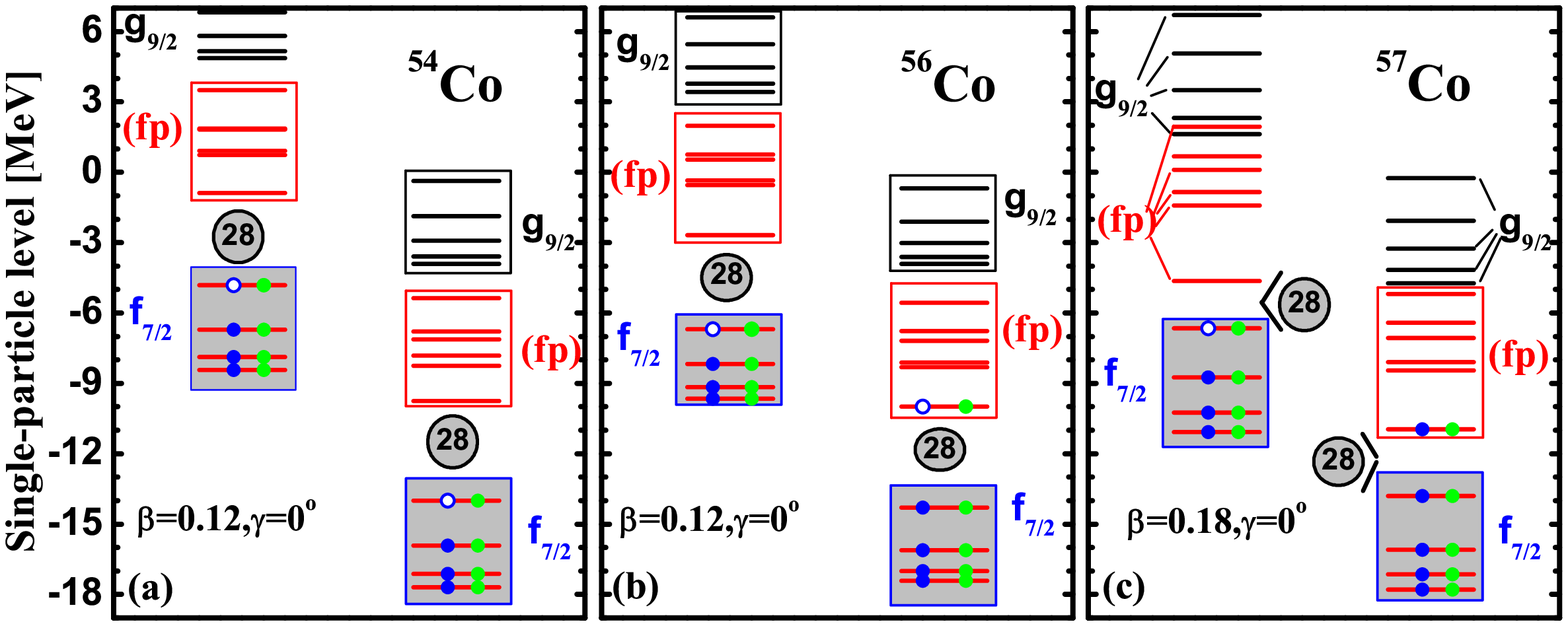}
\includegraphics[width=12 cm]{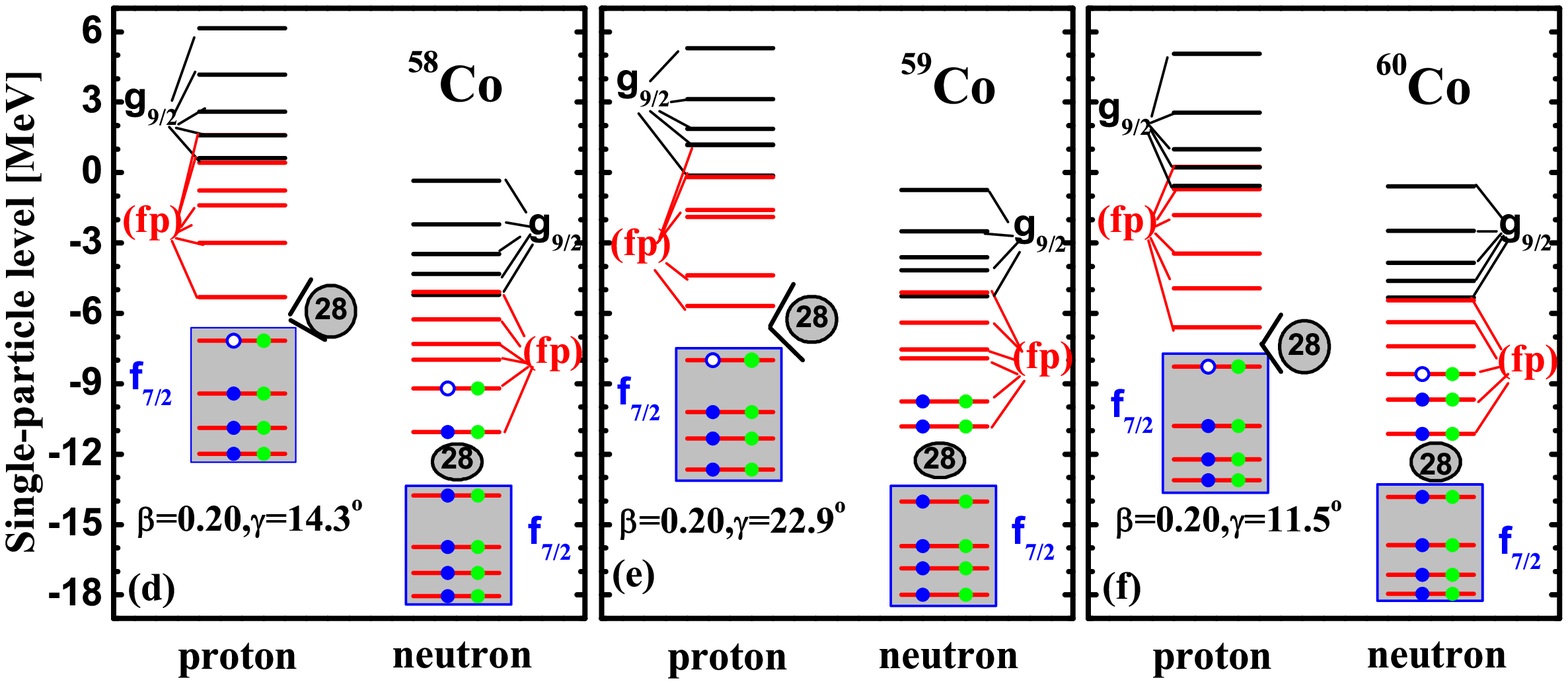}
   \caption{(Color online) Single-proton (left column) and single-neutron (right column)
   Routhians near the Fermi surface in $^{54-60}$Co
              for the ground state.}
   \label{Figocc}
\end{figure*}

Based on adiabatic constrained CDFT calculations, the occupation for
the configuration of the ground states together with the deformation
parameters ($\beta$, $\gamma$) in nuclei $^{54-60}$Co are obtained.
In Fig.~\ref{Figocc}, we show the single-particle energy levels of
protons (left column) and neutrons (right column) near the Fermi
surface for the ground state in nuclei $^{54-60}$Co. For the ground
state, we solve the Dirac equation by filling in each step of the
iteration the proton and neutron levels according to their energies
from the bottom of the well. As shown in Fig.~\ref{Figocc}, for the
proton single-particle energy levels, there is always a hole sitting
on the top of the $f_{7/2}$ shell in the ground states of the nuclei
$^{54-60}$Co due to the lack of a proton with respect to the $Z=28$
full shell. For the neutron single-particle energy levels, there is
also a hole sitting on the top of the $f_{7/2}$ shell in $^{54}$Co,
since its neutron number is identical as the proton number. With the
increase of neutron number, the $f_{7/2}$ shell is filled and the
neutron is filled gradually onto the $(fp)$ shell above $N=28$.
Meanwhile, the deformation parameter $\beta$ increases gradually
from 0.12 at $^{54}$Co to 0.20 at $^{60}$Co, and also the shape
changes from prolate at $^{54,56,57}$Co to triaxial at
$^{58,59,60}$Co. Therefore, for these ground states, the proton has
already played a role of high-$j$ hole, but there is not high-$j$
particle involved. To search for the possible high-$j$ particle-hole
configurations that suitable establish chiral doublet bands, one can
excite the neutron from the $(fp)$ shell to the lowest $g_{9/2}$
orbital. Such kind of the one-particle-one-hole neutron excitation
leads to the valence nucleon configuration with the form of $\pi
f^{-1}_{7/2} \otimes \nu[g^{1}_{9/2}(fp)^n$].

\begin{figure*}
\includegraphics[width=5.4 cm]{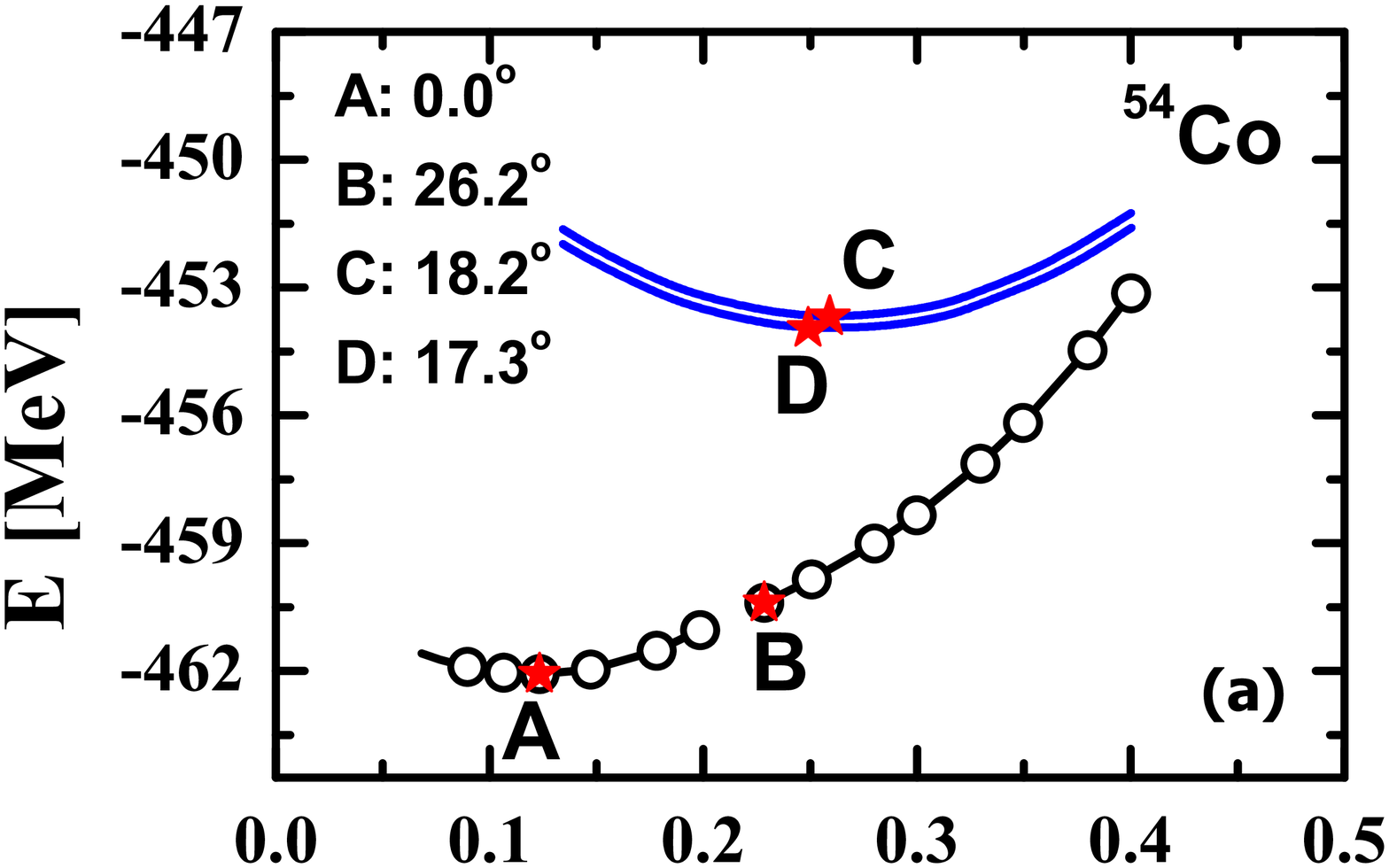}
\includegraphics[width=5.4 cm]{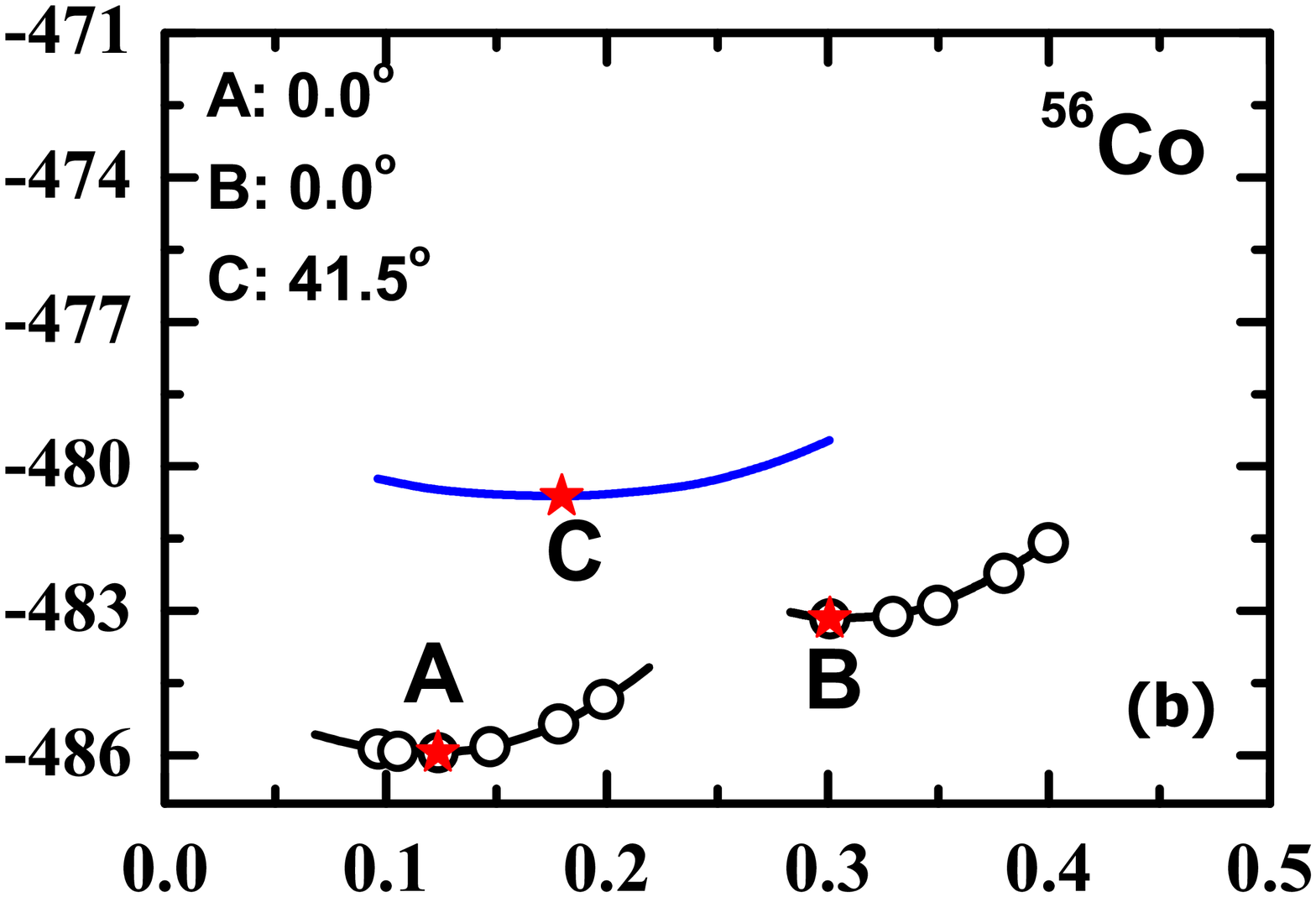}
\includegraphics[width=5.4 cm]{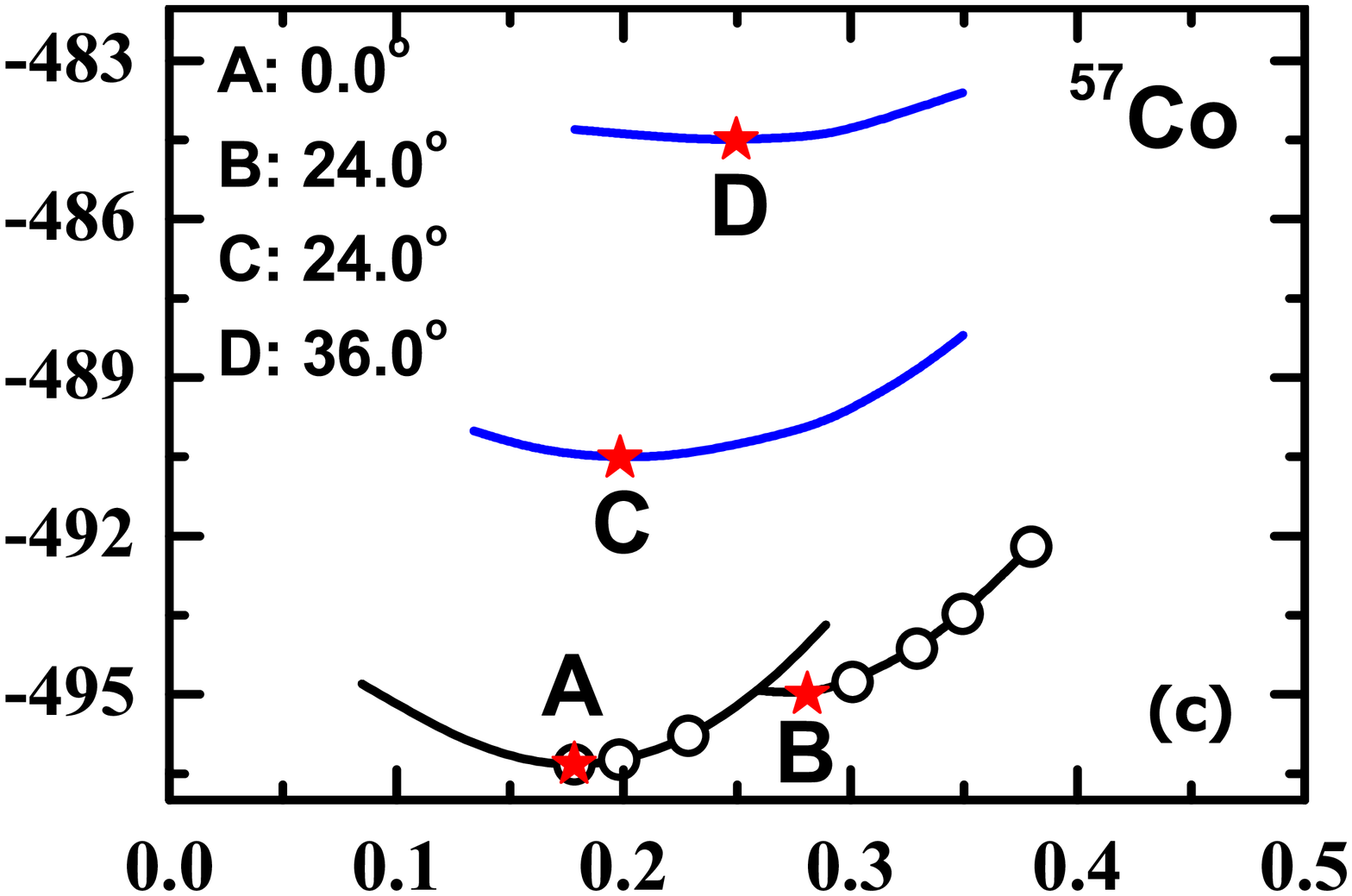}\\
\includegraphics[width=5.4 cm]{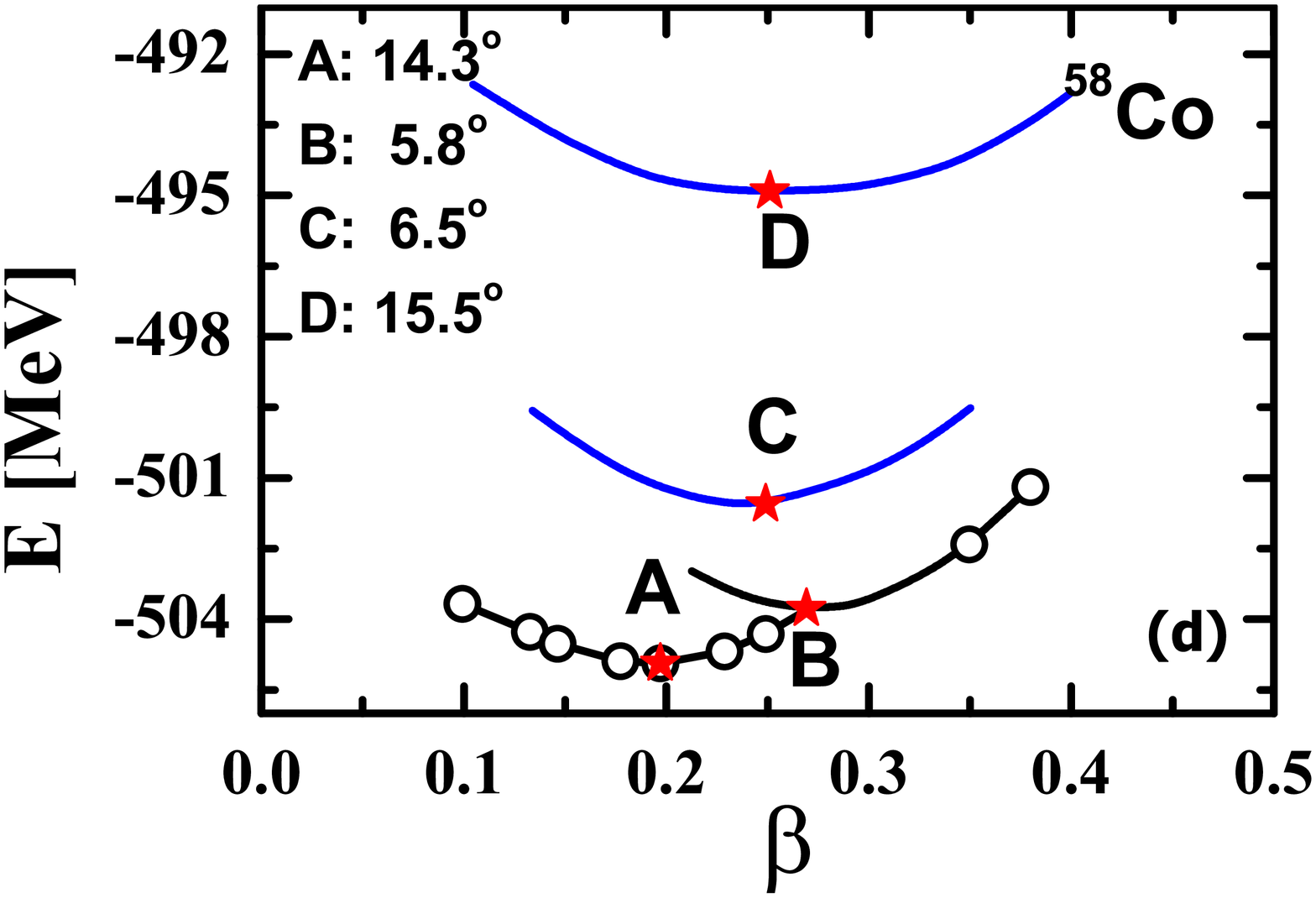}
\includegraphics[width=5.4 cm]{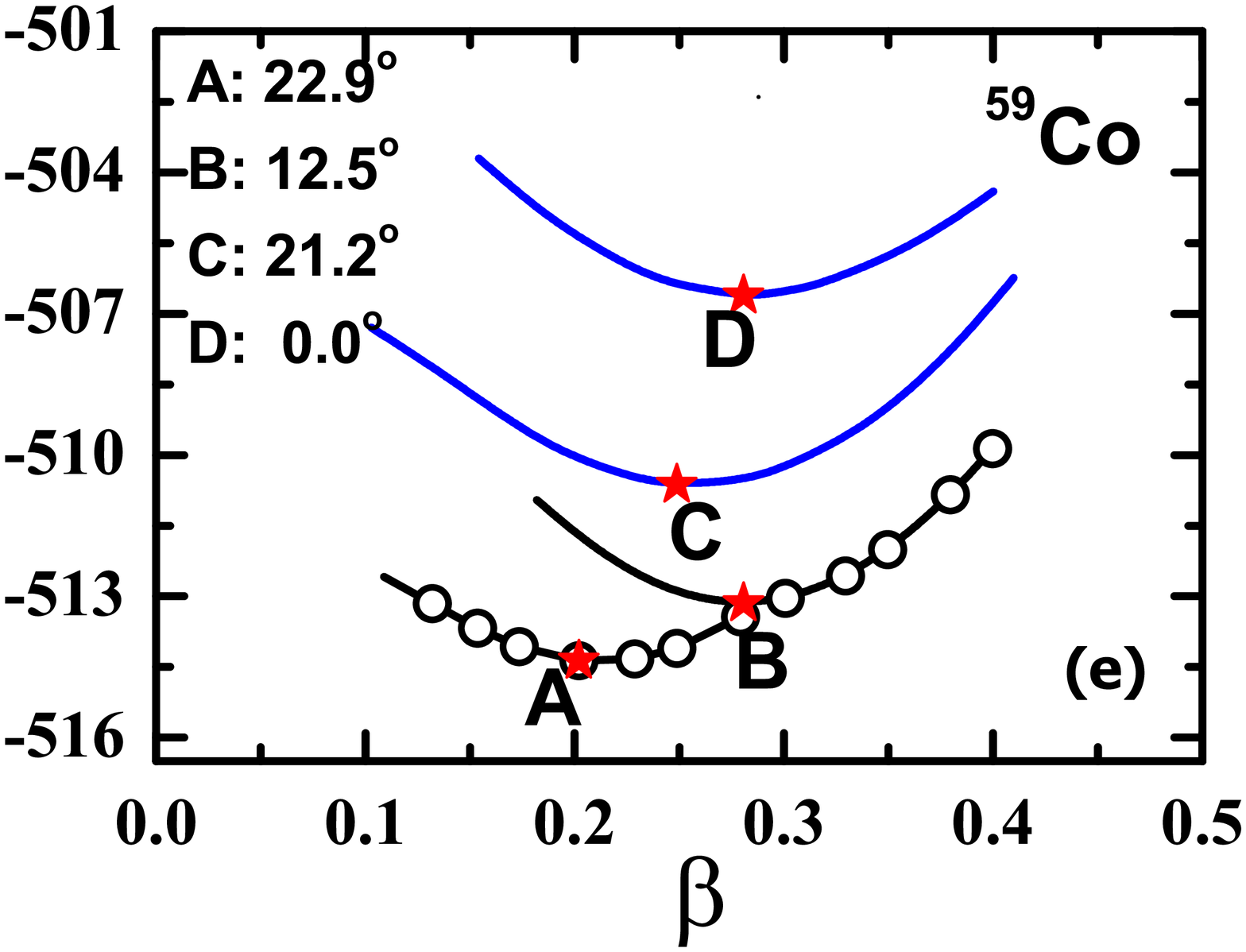}
\includegraphics[width=5.4 cm]{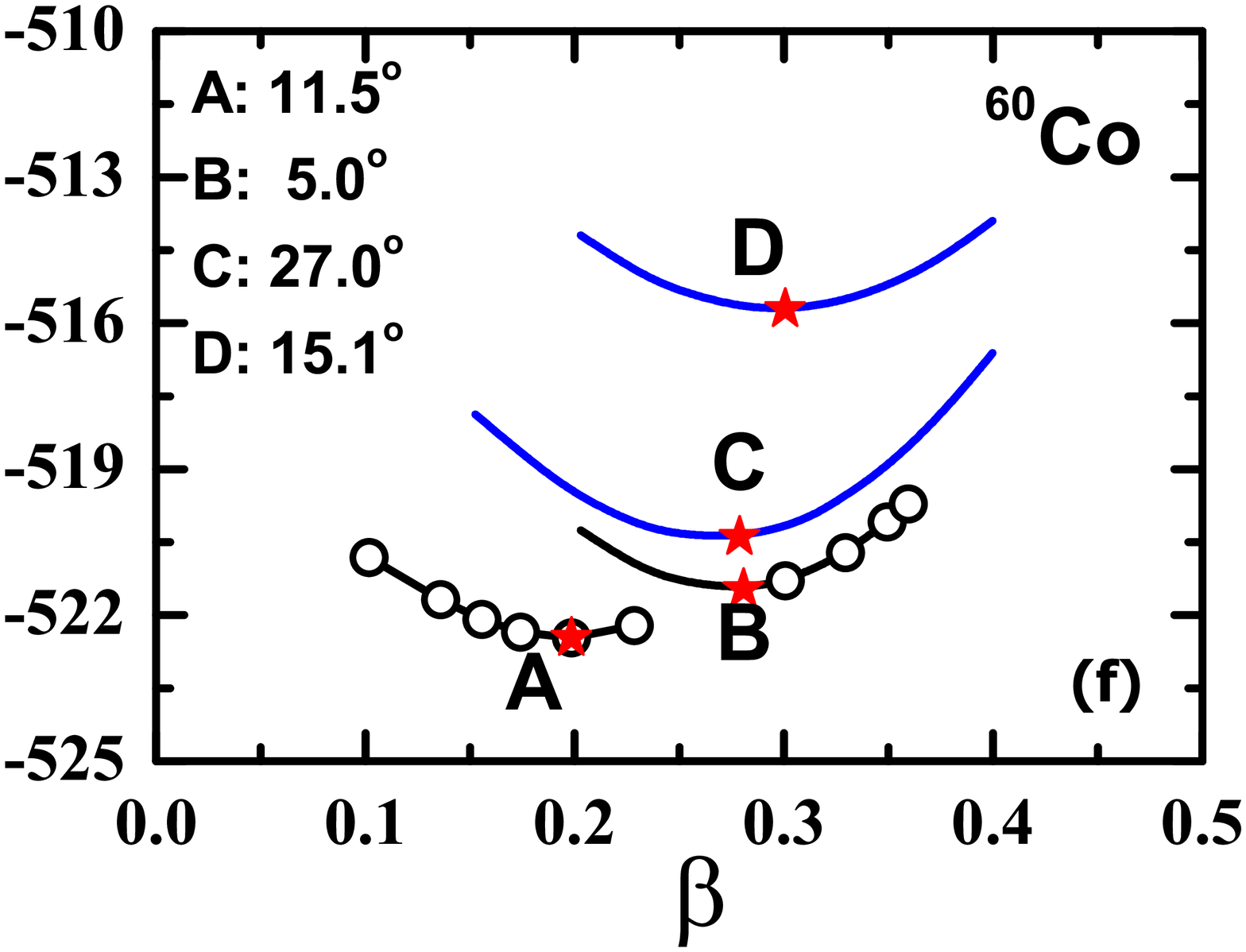}
   \caption{(Color online) The potential-energy curves in adiabatic (open circles)
   and configuration-fixed (solid lines) constrained triaxial CDFT
    calculation with PC-PK1 for $^{54-60}$Co. The local minima in the energy
    surfaces for fixed configuration are represented as stars and labeled
    respectively as A, B, C and D. }
   \label{Figpes}
\end{figure*}

The potential-energy curves for $^{54,56,57,58,59,60}$Co calculated
by adiabatic and configuration-fixed constrained CDFT are presented
as open circles and black solid lines in Fig.~\ref{Figpes},
respectively. In comparison with the irregularities of energy curve
in adiabatic constrained calculations, continuous and smooth energy
curves for each configuration are yielded by the configuration-fixed
constrained calculations. The obvious local minima are represented
by stars and labeled by letters of the alphabet.

Two minima observed in each potential energy curve in
Fig.~\ref{Figpes}, which do not have suitable high-$j$ particle-hole
configurations for chirality, are labeled A and B. Here, state A
represents the ground state, with prolate shape for $^{54,56,57}$Co
and triaxial deformation for $^{58,59,60}$Co. The corresponding
valence nucleon configuration is $\pi f^{-1}_{7/2}\otimes \nu
f^{-1}_{7/2}$ for the ground state of $^{54}$Co and $\pi
f^{-1}_{7/2}\otimes \nu (fp)^{n}$ ($n=1,~2,~3,~4,~5$) for those of
$^{56,57,58,59,60}$Co. The states B in $^{56, 58,59,60}$Co do not
have large enough triaxial deformation parameter. Although states B
of $^{54}$Co and $^{57}$Co have remarkable triaxial deformation,
they do not have proper high-$j$ particle-hole configurations.
Hence, the states A and B are excluded in our waiting list for the
chiral candidate.

By keeping always one neutron at the bottom of the $g_{9/2}$ shell
and the other neutrons filling in the orbitals according to their
energies, low-lying particle-hole excitation states, labeled C in
Fig.~\ref{Figpes}, are obtained. The corresponding configurations
are $\pi f^{-1}_{7/2}\otimes \nu [f^{-2}_{7/2}g^1_{9/2}]$ with the
two $f_{7/2}$ neutrons paired for $^{54}$Co, and $\pi
f^{-1}_{7/2}\otimes \nu [g^1_{9/2}(fp)^n]$ ($n=0,~1,~2,~3,~4$) for
$^{56,57,58,59,60}$Co. Similarly, the configurations $\pi
f^{-1}_{7/2} \otimes \nu[g^{2}_{9/2}(fp)^n]$ ($n=0,~1,~2,~3$) of
$^{57,58,59,60}$Co are connected with a two-particle-two-hole
neutron excitation from the $(fp)$ shell to the two lowest $g_{9/2}$
orbitals ($\nu g_{9/2},m_x=+9/2; \nu g_{9/2},m_x=+7/2$) (states D in
Figs.~\ref{Figpes}(c)-(f)). It should be noted that the mirror
configuration $\pi [f^{-2}_{7/2}g^1_{9/2}]\otimes \nu f^{-1}_{7/2} $
(state D) of state C in $^{54}$Co is obtained by keeping always one
proton at the bottom of the $g_{9/2}$ shell, and the other nucleon
filling the orbitals according to their energies. In the subsequent
calculations with the fixed configurations of C and D, the
occupations of the valence nucleons are traced by the
configuration-fixed constrained
calculations~\cite{Meng2006Phys.Rev.C37303}.

\begin{figure*}
\includegraphics[width=5.4 cm]{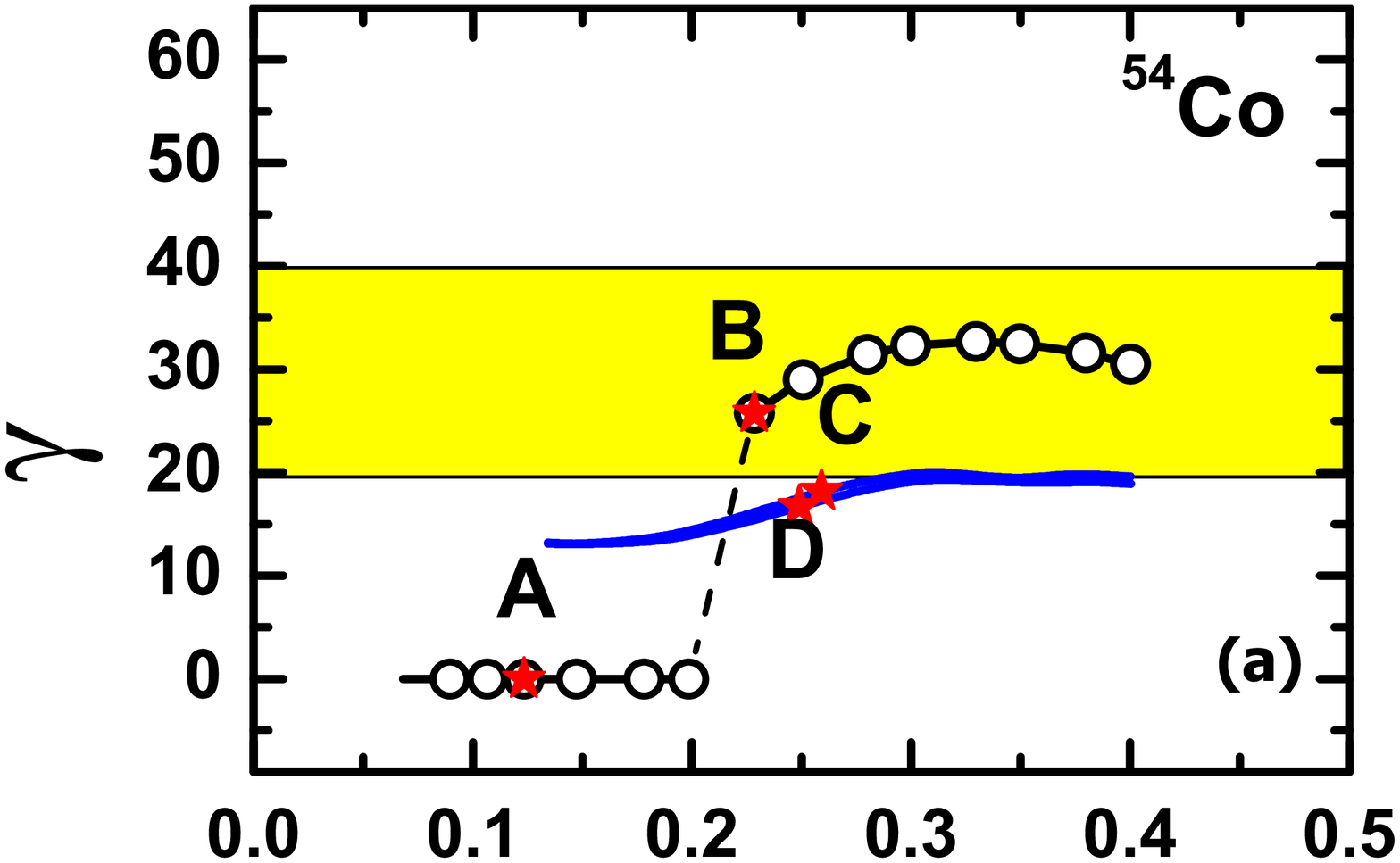}
\includegraphics[width=5.4 cm]{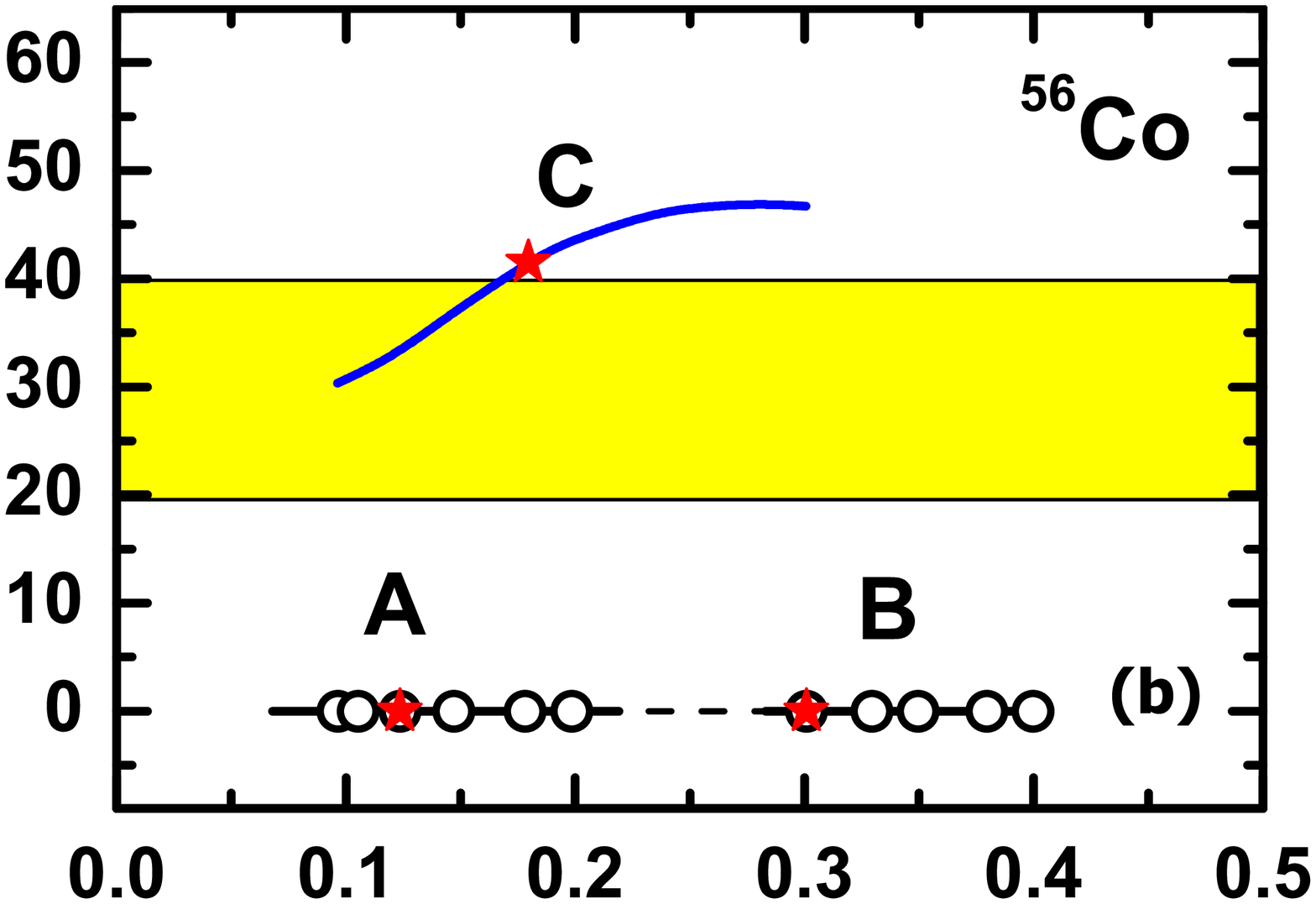}
\includegraphics[width=5.4 cm]{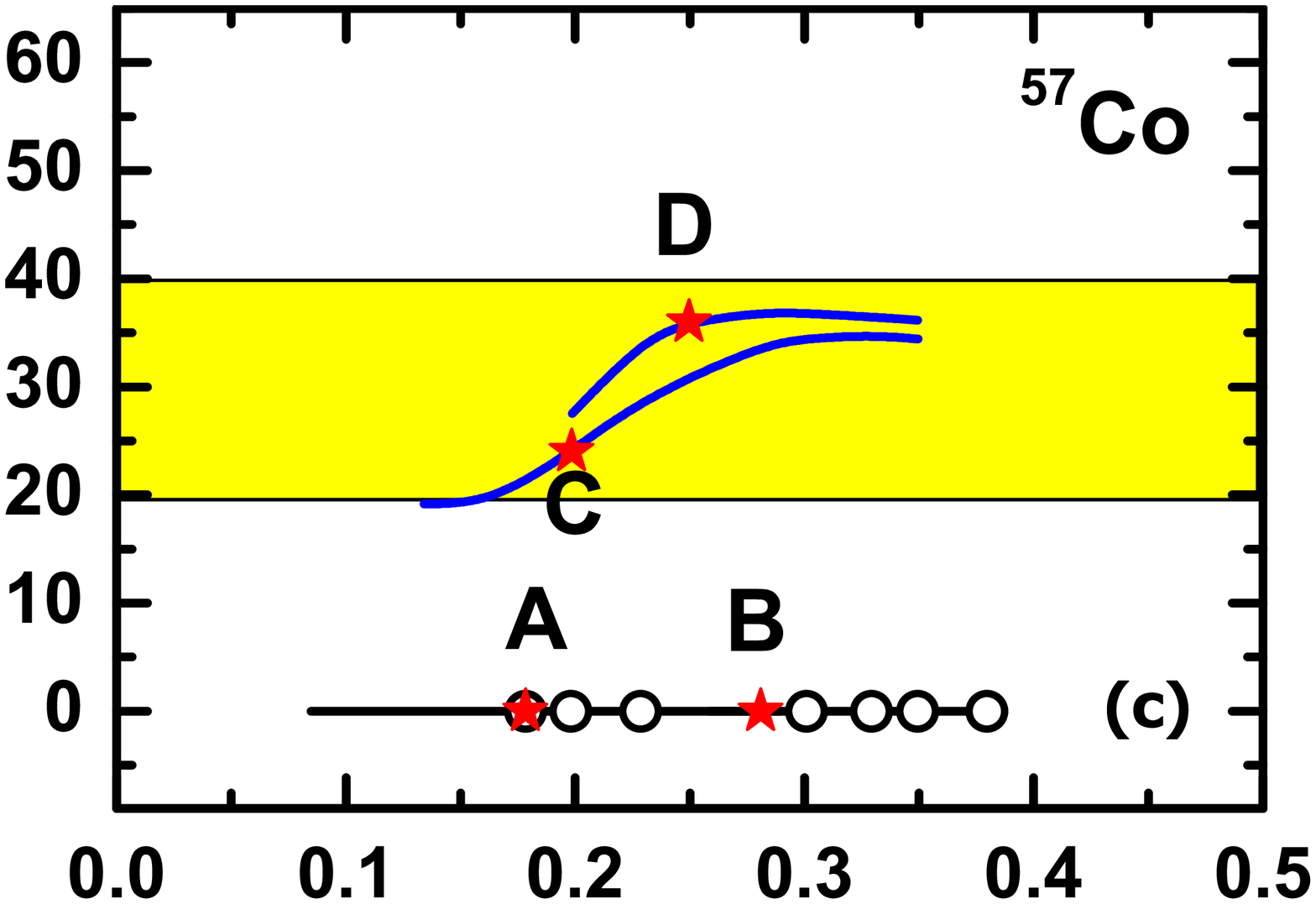}\\
\includegraphics[width=5.4 cm]{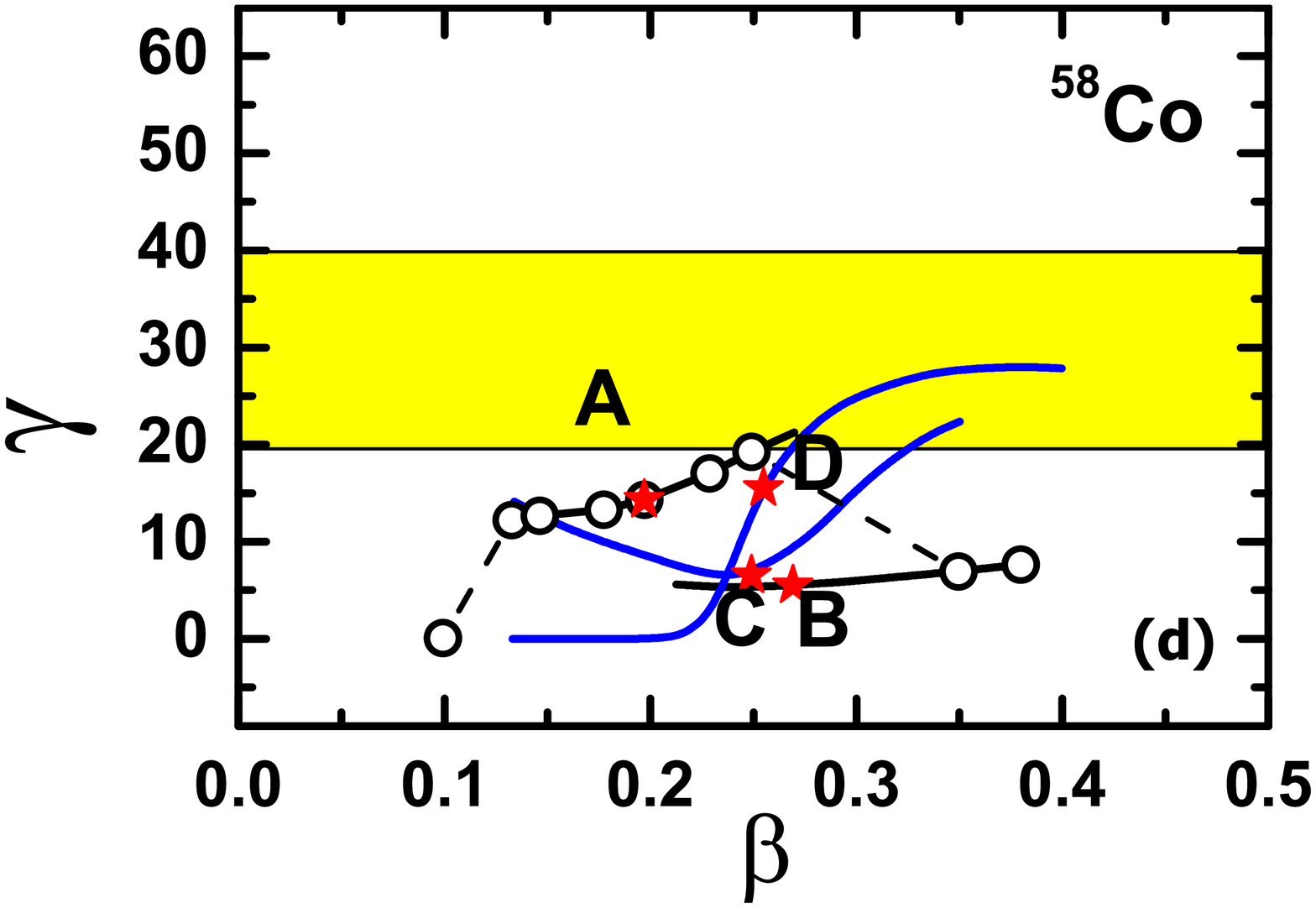}
\includegraphics[width=5.4 cm]{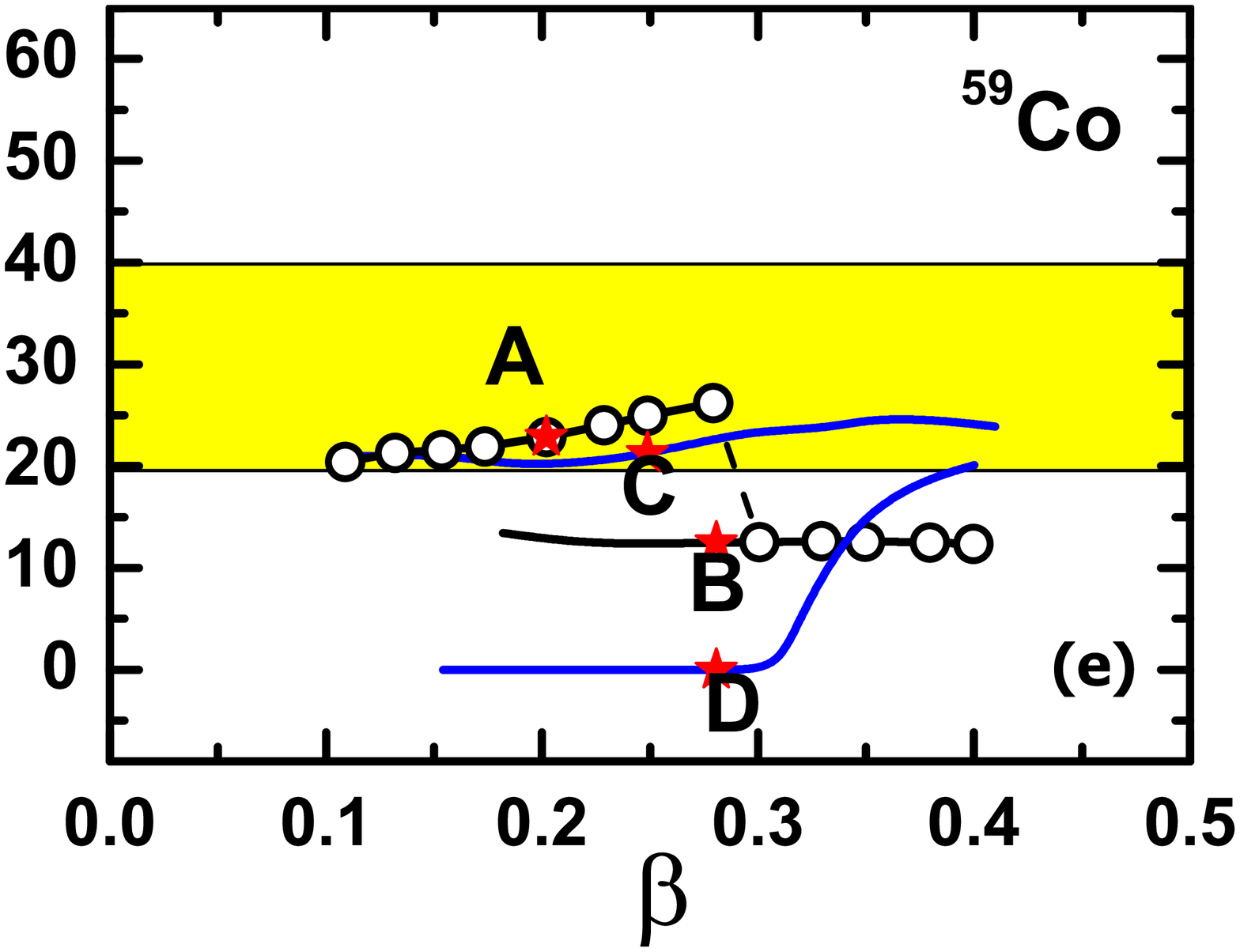}
\includegraphics[width=5.4 cm]{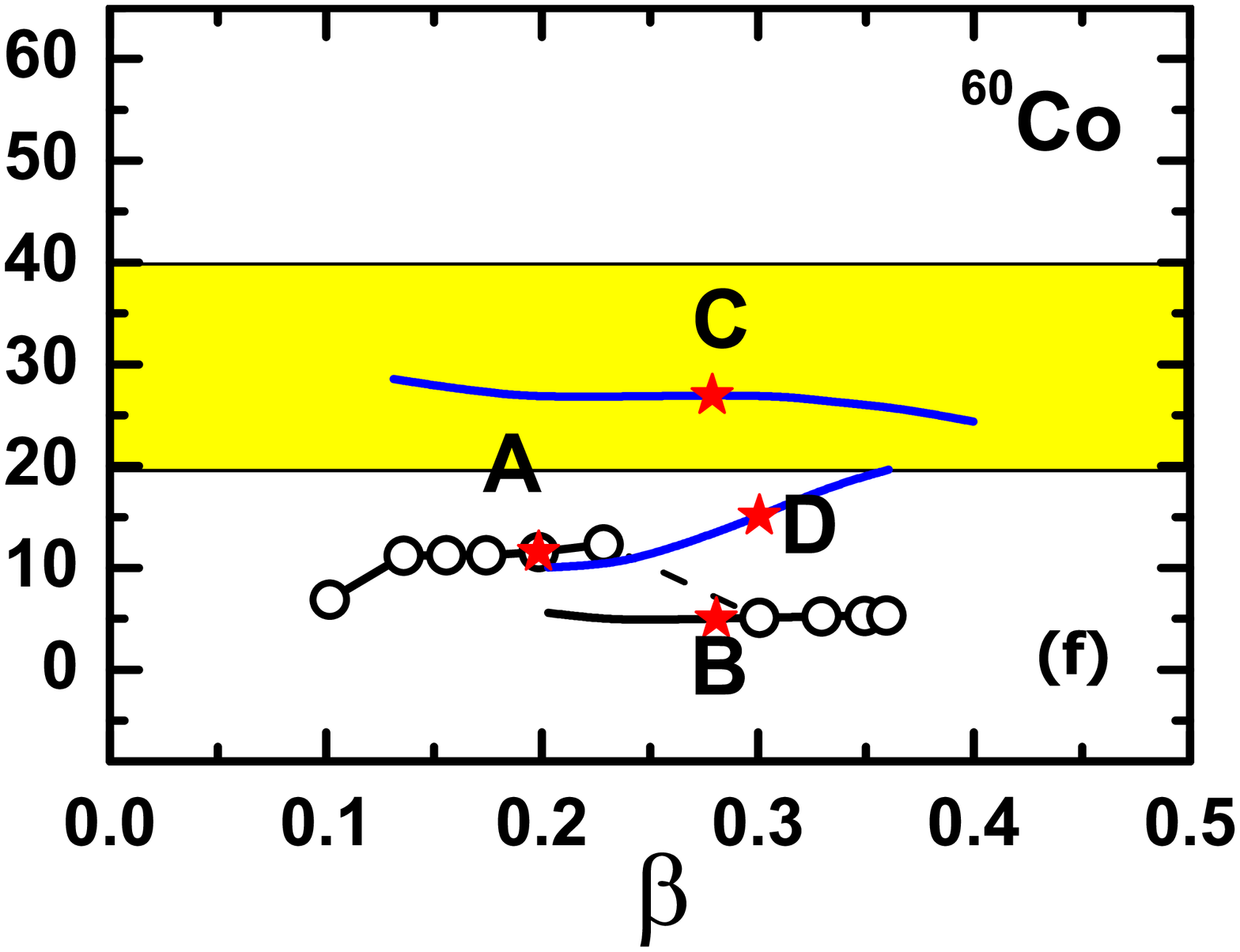}
   \caption{(Color online) The triaxiality parameter $\gamma$ (in degrees) as a function of the deformation
    parameter $\beta$ in adiabatic (open circles) and configuration-fixed (solid lines)
    constrained triaxial CDFT calculations based on the PC-PK1 interaction
    for  $^{54-60}$Co. The local minima in the energy curves for fixed configurations
    are indicated by stars and labeled as  A, B, C and D in accordance with the increasing
    of energy respectively. The shaded area represents the triaxiality parameter $\gamma$
    favorable for nuclear chirality.}
   \label{Figgam}
 \end{figure*}

The obtained results are presented as blue solid lines in
Figs.~\ref{Figpes}(a)-(f). In Fig.~\ref{Figpes}(a), we find that the
potential energy curve with configuration C is similar to that with
configuration D for $^{54}$Co. Both minima C and D in $^{54}$Co have
deformation parameters $\beta$ and $\gamma$ suitable for chirality,
which are C($\beta=0.259$, $\gamma=18.2^\circ$) and D($\beta=0.259$,
$\gamma=17.3^\circ$). In addition to the states C and D in
$^{54}$Co, there are also several excited local minima with
prominent triaxial deformations in $^{56,57,58,59,60}$Co. These
triaxial local minima are state C($\beta=0.180$,
$\gamma=41.5^\circ$) for $^{56}$Co, states C($\beta=0.199$,
$\gamma=24.0^\circ$) and D($\beta=0.250$, $\gamma=36.0^\circ$) for
$^{57}$Co, state D($\beta=0.251$, $\gamma=15.5^\circ$) for
$^{58}$Co, state C($\beta=0.249$, $\gamma=21.2^\circ$) for
$^{59}$Co, states C($\beta=0.279$, $\gamma=27.0^\circ$) and
D($\beta=0.301$, $\gamma=15.1^\circ$) for $^{60}$Co. All these
states have triaxial deformations as well as high-$j$ particle-hole
configurations that suitable for establishing chiral rotation.
Therefore, the existences of chiral doublets or M$\chi$D can be
expected in $^{54,56,57,58,59,60}$Co.

\begin{figure*}
\includegraphics[width=7.8 cm]{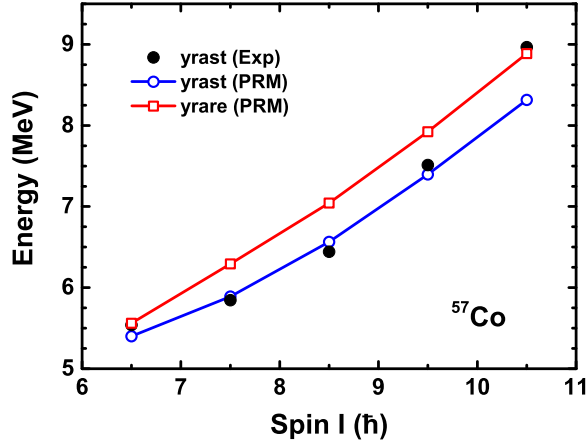}
   \caption{(Color online) The calculated energy spectra by particle rotor model in comparison with the experimental data in Ref.~\cite{Caballero2003PRC}. More details can be seen in the text.}
   \label{Co57_PRM}
 \end{figure*}

It is worth mentioning that in
Ref.~\cite{Caballero2003PRC}, a dipole band with the configuration
$\pi f^{-1}_{7/2}\otimes\nu  [g^{1}_{9/2}(fp)^1]$ was already observed
in $^{57}$Co. It is built on the $I=13/2^+$ state with band-head excitation energy 5.539 MeV~\cite{Caballero2003PRC}. In the CDFT calculation, the excitation energy of the configuration is 5.790 MeV, which agree reasonably with the experimental value. With the configuration and the corresponding deformation parameters ($\beta=0.199$, $\gamma=24^\circ$), the PRM~\cite{Ayangeakaa2013PRL, Kuti2014PRL, Lieder2014PRL, Q.B.Chen2018PLB744} calculation is performed and the obtained energy spectra in comparison with the experimental data (cascade 1 in Ref.~\cite{Caballero2003PRC}) is shown in Fig.~\ref{Co57_PRM}. In the calculation, the moments of inertia are taken as $\mathcal{J}_k=\mathcal{J}_0 \sin^2 (\gamma-2k\pi/3)$ with $\mathcal{J}_0=6.0~\hbar^2/\textrm{MeV}$ and the Coriolis attenuation factor is taken as $\xi=0.9$. It can be seen that the calculated results can reproduce reasonably the experimental yrast band, except for the data point at $I=10.5\hbar$ (which was tentative assigned in Ref.~\cite{Caballero2003PRC}). The energy difference between the calculated partner bands is about 400 keV. This value could correspond to the picture of chiral vibration motion in the low spin region of chiral doublets. Hence, it is encouraged to further search for the partner band in this nucleus from experiment.


\begin{table}[h!tbp]
 \centering\setlength{\tabcolsep}{0.35em}\fontsize{10pt}{11pt}\selectfont
 \caption{Energies (in MeV), deformation parameters $\beta$ and $\gamma$ (in degree), as
         well as the corresponding configurations (both valence nucleon and unpaired
         nucleon) of the local minima in $^{54-60}$Co obtained by the configuration-fixed
         constrained triaxial CDFT calculations.}
\begin{tabular}{ c c c c c c c}
 \hline
\hline
 \multicolumn{1}{ c }{{\hspace{0.1cm}}{\hspace{0.1cm}}}
 &\multicolumn{1}{c }{{\hspace{0.1cm}} State {\hspace{0.1cm}} }
  &\multicolumn{1}{c@{\hspace{0.5cm}} }{ ~~~~~~~~Valence-cfg. }
  &~~~~~~~~~~~Unpaired-cfg.~~~~~
  & ~~~Energy~~   &  $\beta$
  &\multicolumn{1}{c@{\hspace{1.0cm}} }{ ~~$\gamma$ } \\
\cline{1-7}
\multirow{4}{*}{$^{54}$Co}
 &A & $\pi f^{-1}_{7/2} \otimes \nu f^{-1}_{7/2}$
    &~$\pi f^{-1}_{7/2} \otimes \nu f^{-1}_{7/2}$
    & $-462.07$ & 0.123 & ~~0.0$^\circ$ \\
\cline{2-7}
 &B & $\pi [f^{-2}_{7/2}(fp)^{1}]\otimes\nu f^{-1}_{7/2}$
    &~$\pi (fp)^1 \otimes \nu f^{-1}_{7/2}$
    & $-460.34$ &0.231 &26.2$^\circ$ \\
\cline{2-7}
 &C & $\pi f^{-1}_{7/2} \otimes \nu [f^{-2}_{7/2}g^{1}_{9/2}]$
    & $\pi f^{-1}_{7/2} \otimes \nu g^{1}_{9/2}$
    & $-453.68$ &0.259 &18.2$^\circ$ \\
\cline{2-7}
 &D & $\pi [f^{-2}_{7/2}g^{1}_{9/2}] \otimes \nu f^{-1}_{7/2}$
    & $\pi g^{1}_{9/2} \otimes \nu f^{-1}_{7/2}$
    & $-453.97$ &0.259 &17.3$^\circ$ \\
\hline
\multirow{3}{*}{$^{56}$Co}
 &A & $\pi f^{-1}_{7/2} \otimes \nu (fp)^1$
    &~$\pi f^{-1}_{7/2} \otimes \nu (fp)^1$
    & $-485.95$ & 0.124 & ~~0.0$^\circ$ \\
\cline{2-7}
 &B & $\pi [f^{-2}_{7/2}(fp)^{1}]\otimes\nu [f^{-1}_{7/2}(fp)^2]$
    &~$\pi (fp)^{1}\otimes \nu f^{-1}_{7/2}$
    & $-483.17$ &0.301 &~~0.0$^\circ$ \\
\cline{2-7}
 &C & $\pi f^{-1}_{7/2}\otimes\nu g^{1}_{9/2}$
    &~$\pi f^{-1}_{7/2}\otimes\nu g^{1}_{9/2}$
    & $-480.61$ &0.180 &41.5$^\circ$ \\
\hline
\multirow{4}{*}{$^{57}$Co}
 &A & $\pi f^{-1}_{7/2} \otimes \nu (fp)^2$
    & $\pi f^{-1}_{7/2}$
    & $-496.32$ & 0.178 &   ~~0.0$^\circ$ \\
\cline{2-7}
 &B & $\pi [f^{-2}_{7/2}(fp)^{1}]\otimes\nu (fp)^{2}$
    & $\pi (fp)^{1}$
    & $-495.00$ & 0.281 &24.0$^\circ$ \\
\cline{2-7}
 &C & $\pi f^{-1}_{7/2}\otimes\nu [g^{1}_{9/2}(fp)^1]$
    & $\pi f^{-1}_{7/2}\otimes\nu  [g^{1}_{9/2}(fp)^1]$
    & $-490.53$ & 0.199 &24.0$^\circ$ \\
\cline{2-7}
 &D & $\pi f^{-1}_{7/2}\otimes\nu g^{2}_{9/2}$
    & $\pi f^{-1}_{7/2}\otimes\nu g^{2}_{9/2}$
    & $-484.50$ & 0.250 & 36.0$^\circ$ \\
\hline
\multirow{4}{*}{$^{58}$Co}
 &A & $\pi f^{-1}_{7/2} \otimes \nu (fp)^3$
    & $\pi f^{-1}_{7/2} \otimes \nu (fp)^1$
    & $-504.94$&0.197& 14.3$^\circ$ \\
\cline{2-7}
 &B & $\pi [f^{-2}_{7/2}(fp)^{1}]\otimes\nu (fp)^3$
    & $\pi (fp)^{1}\otimes\nu (fp)^1$
    & $-503.77$ & 0.269 & ~~5.5$^\circ$ \\
\cline{2-7}
 &C & $\pi f^{-1}_{7/2}\otimes\nu [g^{1}_{9/2}(fp)^{2}]$
    & $\pi f^{-1}_{7/2}\otimes\nu g^{1}_{9/2}$
    & $-501.56$ & 0.249 & ~~6.5$^\circ$ \\
\cline{2-7}
 &D & $\pi f^{-1}_{7/2}\otimes\nu [g^{2}_{9/2}(fp)^{1}]$
    & $\pi f^{-1}_{7/2}\otimes\nu [g^{2}_{9/2}(fp)^{1}]$
    & $-494.91$&0.251& 15.5$^\circ$ \\
\hline
\multirow{4}{*}{$^{59}$Co}
 &A & $\pi f^{-1}_{7/2} \otimes \nu (fp)^4$
    & $\pi f^{-1}_{7/2}$
    & $-514.36$&0.202& 22.9$^\circ$ \\
\cline{2-7}
 &B & $\pi [f^{-2}_{7/2}(fp)^{1}]\otimes\nu (fp)^4$
    & $\pi (fp)^{1}$
    & $-513.15$ & 0.281 & 12.5$^\circ$ \\
\cline{2-7}
 &C & $\pi f^{-1}_{7/2}\otimes\nu [g^{1}_{9/2}(fp)^{3}]$
    & $\pi f^{-1}_{7/2}\otimes\nu [g^{1}_{9/2}(fp)^{1}]$
    & $-510.63$ & 0.249 & 21.2$^\circ$ \\
\cline{2-7}
 &D & $\pi f^{-1}_{7/2}\otimes\nu [g^{2}_{9/2}(fp)^{2}]$
    & $\pi f^{-1}_{7/2}\otimes\nu g^{2}_{9/2}$
    & $-506.61$&0.281& ~~0.0$^\circ$ \\
\hline
\multirow{4}{*}{$^{60}$Co}
 &A & $\pi f^{-1}_{7/2} \otimes \nu (fp)^5$
    & $\pi f^{-1}_{7/2}\otimes\nu (fp)^{1}$
    & $-522.46$&0.199& 11.5$^\circ$ \\
\cline{2-7}
 &B & $\pi [f^{-2}_{7/2}(fp)^{1}]\otimes\nu (fp)^5$
    & $\pi (fp)^{1}\otimes\nu (fp)^{1}$
    & $-521.46$ & 0.270 & ~~5.0$^\circ$ \\
\cline{2-7}
 &C & $\pi f^{-1}_{7/2}\otimes\nu [g^{1}_{9/2}(fp)^{4}]$
    & $\pi f^{-1}_{7/2}\otimes\nu g^{1}_{9/2}$
    & $-520.39$ & 0.279 & 27.0$^\circ$ \\
\cline{2-7}
 &D & $\pi f^{-1}_{7/2}\otimes\nu [g^{2}_{9/2}(fp)^{3}]$
    & $\pi f^{-1}_{7/2}\otimes\nu [g^{2}_{9/2}(fp)^{1}]$
    & $-515.71$&0.301& 15.1$^\circ$ \\
\hline
\hline
\end{tabular}
 \label{tab:1}
\end{table}

The triaxial deformation parameter $\gamma$ of
$^{54,56,57,58,59,60}$Co as functions of deformation $\beta$ in
adiabatic and configuration-fixed constrained CDFT calculations are
presented as open circles and solid lines in Fig.~\ref{Figgam},
respectively. In each panel, the stars labeled by letter of alphabet
correspond to the deformation of the local minima in
Fig.~\ref{Figpes}, and the yellow area represents favorable triaxial
deformation for chirality. In the $\beta$-$\gamma$ curves calculated
by the adiabatic constrained calculations, the jumps marked by the
dashed-lines represent energy-favorable configurations change. In
contrast, the smooth $\beta$-$\gamma$ curves for the given
configurations are obtained by the configuration-fixed constrained
calculations. It should be noted that triaxiality in
$^{54,56,57,58,59,60}$Co favorable for nuclear chirality appears
roughly at $\beta = 0.1 - 0.3$.


The energies, deformation parameters $\beta$ and $\gamma$
as well as the corresponding valence nucleon and unpaired nucleon
configurations for the local minima in $^{54,56,57,58,59,60}$Co,
obtained by the configuration-fixed constrained triaxial CDFT
calculations, are summarized in Table~\ref{tab:1}. Although there
are no high-$j$ neutron particles for the ground state in
$^{54,56,57,58,59,60}$Co, we find that the high-$j$ particle and
hole configurations are available by single-particle excitations,
such as for the triaxially deformed local minima C and D in
$^{54}$Co, C in $^{56}$Co, C and D in $^{57}$Co, D in $^{58}$Co, C
in $^{59}$Co, and C and D in $^{60}$Co. Therefore, all of these
nuclei could possibly exhibit chiral rotation characters based on
the above mentioned configurations. In particular, in $^{54}$Co,
$^{57}$Co and $^{60}$Co, there are two triaxially deformed states
with high-$j$ particle and hole configurations. Further experimental
efforts to search for M$\chi$D in these cobalt isotopes are highly
encouraged.

\begin{figure*}
\includegraphics[width=7.8 cm]{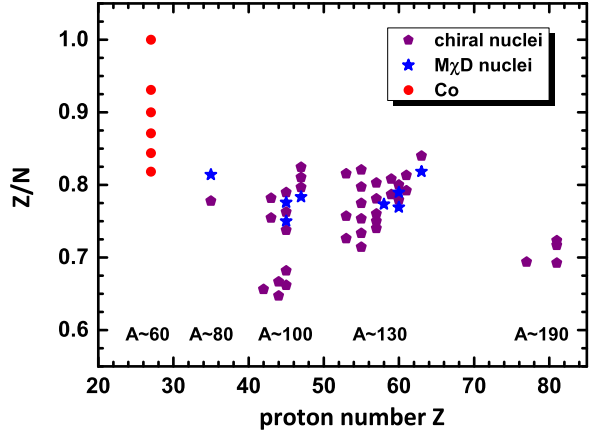}
   \caption{(Color online) The ratio $Z/N$ of proton number and neutron number
   of Co isotopes are compared with those of reported chiral nuclei and M$\chi$D nuclei in
   the $A\sim 80$, $100$, $130$, and $190$ mass regions. The data are from
   Ref.~\cite{Xiong2018arXiv} and references therein.}
   \label{Z_N_ratio}
 \end{figure*}

In Fig.~\ref{Z_N_ratio}, the ratio $Z/N$ of proton number and
neutron number of Co isotopes are compared with those of reported
chiral nuclei and M$\chi$D nuclei in the $A\sim80, 100, 130,$ and
$190$ mass regions. The data are from Ref.~\cite{Xiong2018arXiv} and
references therein. It can be found that the $Z/N$ ratios of the
reported chiral and M$\chi$D nuclei in the $A\sim80, 100, 130,$ and
$190$ mass regions locate around the region from 0.65 to 0.85. For
the Co isotopes, the ratio is from $\sim 0.80$ in $^{60}$Co to 1.00
in $^{54}$Co. This is a unique feature in the Co isotopes, and
provides the possibility that existence of M$\chi$D with mirror
configuration as in $^{54}$Co. If the M$\chi$D was successfully
identified in the experiment, it provides a promising example to
investigate the neutron proton interactions in the chiral doublet
bands.


In summary, the possible existence of chiral or multiple chiral
doublets (M$\chi$D) in cobalt isotopes are investigated by the
adiabatic and configuration-fixed constrained CDFT approaches. The
potential-energy curves and triaxial deformation parameters $\gamma$
as functions of the deformation parameter $\beta$ in
$^{54,56,57,58,59,60}$Co are obtained. Serval local energy minima
with prominent triaxial deformations are found in these cobalt
isotopes. According to the transformed quantum numbers
$\left<{nljm}\right>$, the configurations of these triaxially
deformed minima are specified. It is found that the configurations
$\pi f^{-1}_{7/2} \otimes \nu[g^{m}_{9/2}(fp)^n$] ($m$ = 1 or 2) are
available for the triaxially deformed local minima in all of these
isotopes. For $^{54}$Co, there are mirror configurations with $\pi
[f^{-2}_{7/2}g^1_{9/2}]\otimes \nu f^{-1}_{7/2} $ and $\pi
f_{7/2}^{-1}\otimes \nu [f_{7/2}^{-2}g_{9/2}^1]$, and both states
are of the triaxiality. Based on the triaxial deformation and the
corresponding high-$j$ particle and hole configurations, there could
exist chiral rotation characters in $^{54,56,57,58,59,60}$Co and
M$\chi$D in $^{54, 57, 60}$Co. Hence, the present investigation
provides not only the prediction of chirality and M$\chi$D in
$A\sim60$ mass region, but also presents an experimental opportunity
for the observation of chirality in this mass region.

\vspace{2em} Helpful discussions with  J. Meng and S. Q. Zhang are
gratefully acknowledged. This work is supported by the National
Natural Science Foundation of China (NSFC) under Grant No. 11775026,
the Open Project Program of State Key Laboratory of Theoretical
Physics, Institute of Theoretical Physics, Chinese Academy of
Sciences, China (No. Y4KF041CJ1), and the Deutsche
Forschungsgemeinschaft (DFG) and NSFC through funds provided to the
Sino-German CRC 110 ``Symmetries and the Emergence of Structure in
QCD''.

\end{document}